\documentclass[12pt]{article}
\usepackage{amsmath}
\usepackage{graphicx,psfrag,epsf}
\usepackage{enumerate}
\usepackage{natbib}
\usepackage{textcomp}
\usepackage[hyphens]{url} 
\usepackage{hyperref}
\providecommand{\tightlist}{%
  \setlength{\itemsep}{0pt}\setlength{\parskip}{0pt}}

\newcommand{\blind}{0}

\usepackage{xcolor}
\xdefinecolor{blue}{rgb}{0.28,0.65,0.9}
\hypersetup{
    colorlinks,
    linkcolor={blue},
    citecolor={blue},
    urlcolor={blue}
}

\usepackage{booktabs}

\usepackage{amssymb}
\usepackage[T1]{fontenc}

\newcommand{\Beckman}[1]{Beckman}
\newcommand{\Tackett}[1]{Tackett}
\newcommand{\Rundel}[1]{Rundel}
\newcommand{\Sullivan}[1]{Sullivan}

\newcommand{\BeckmanBlind}[1]{Inst.\ A}
\newcommand{\TackettBlind}[1]{Inst.\ B}
\newcommand{\RundelBlind}[1]{Inst.\ C}
\newcommand{\SullivanBlind}[1]{Inst.\ D}
\usepackage{booktabs}
\usepackage{longtable}
\usepackage{array}
\usepackage{multirow}
\usepackage{wrapfig}
\usepackage{float}
\usepackage{colortbl}
\usepackage{pdflscape}
\usepackage{tabu}
\usepackage{threeparttable}
\usepackage{threeparttablex}
\usepackage[normalem]{ulem}
\usepackage{makecell}
\usepackage{xcolor}

\begin{document}

\def\spacingset#1{\renewcommand{\baselinestretch}%
{#1}\small\normalsize} \spacingset{1}


\if0\blind
{
  \title{\bf Implementing version control with Git and GitHub as a
learning objective in statistics and data science courses}

  \author{
        Matthew D. Beckman \\
    Penn State University\\
     and \\     Mine \c{C}etinkaya-Rundel \\
    University of Edinburgh, RStudio, Duke University\\
     and \\     Nicholas J. Horton \\
    Amherst College\\
     and \\     Colin W. Rundel \\
    University of Edinburgh, Duke University\\
     and \\     Adam J. Sullivan \\
    Brown University\\
     and \\     Maria Tackett \\
    Duke University\\
      }
  \maketitle
} \fi

\if1\blind
{
  \bigskip
  \bigskip
  \bigskip
  \begin{center}
    {\LARGE\bf Implementing version control with Git and GitHub as a
learning objective in statistics and data science courses}
  \end{center}
  \medskip
} \fi

\bigskip
\begin{abstract}
A version control system records changes to a file or set of files over
time so that changes can be tracked and specific versions of a file can
be recalled later. As such, it is an essential element of a reproducible
workflow that deserves due consideration among the learning objectives
of statistics courses. This paper describes experiences and
implementation decisions of four contributing faculty who are teaching
different courses at a variety of institutions. Each of these faculty
have set version control as a learning objective and successfully
integrated one such system (Git) into one or more statistics courses.
The various approaches described in the paper span different
implementation strategies to suit student background, course type,
software choices, and assessment practices. By presenting a wide range
of approaches to teaching Git, the paper aims to serve as a resource for
statistics and data science instructors teaching courses at any level
within an undergraduate or graduate curriculum.

In press, \emph{Journal of Statistics and Data Science Education}
\end{abstract}

\noindent%
{\it Keywords:} statistical computing, education, data acumen, data
science, reproducible analysis, workflow, collaborative learning
\vfill

\newpage
\spacingset{1.45} 

\hypertarget{introduction}{%
\section{Introduction}\label{introduction}}

\citet{nolan_templelang_2010} promote ``version control'' as a key topic
for statistical analysis, particularly when coordinating work across a
team. A version control system records changes to a file or set of files
over time so that changes can be tracked and specific versions of a file
can be recalled later.

The \emph{2014 American Statistical Association Curriculum Guidelines
for Undergraduate Programs} includes proficiency with modern statistical
software as well as well-documented and reproducible data wrangling
skills as a necessary component of the undergraduate statistics
curriculum \citep{asa_guidelines_2014}. The National Academies consensus
report on \emph{Data Science for Undergraduates} \citep{nasem_2018}
identifies workflow and reproducibility as important components of
``data acumen''. Version control is an important foundation for
reproducible workflows, be they collaborative (maintaining versions of
files that are being modified by teams) or non-collaborative (tracking
analysis histories and providing analysis provenance). It forms a
necessary part of a reproducible workflow, and therefore deserves due
consideration among the learning objectives of statistics and data
science courses. \citet{fiksel_2019} motivate the use of GitHub for
version control and describe how they integrated this complex and
powerful system into two courses.

This paper follows a similar format to \citet{garfield_2011} and
\citet{hardin_2015} by describing the experiences and implementation
decisions of several contributing faculty---teaching different courses
at different institutions---who have successfully integrated Git into
one or more statistics courses to teach version control as a learning
objective. We begin by discussing our motivations for identifying
version control as a learning objective and then provide summaries of
courses taught by the four contributing faculty highlighting different
implementation strategies chosen based on student audience, course type,
software choices, and assessment practices. We highlight a range of
implementations across a variety of courses and student populations in
order to provide a resource for statistics instructors to interpolate an
implementation suitable for use in their own courses at the
undergraduate or graduate level. We refer the reader to Table
\ref{tab:def-table} for definitions of terms we will use regularly
throughout the paper. Readers who are unfamiliar with version control
would benefit from reading \citet{bryan_2018_excuse}.

\begin{longtable}[t]{>{\raggedright\arraybackslash}p{4cm}|>{\raggedright\arraybackslash}p{12cm}}
\caption{\label{tab:def-table}Definitions of common terms.}\\
\toprule
Term & Definition\\
\midrule
\cellcolor{gray!6}{Git} & \cellcolor{gray!6}{An open source version control software system (\href{https://git-scm.com/}{git-scm.com})}\\
Git repository (or repo) & Analogous to a project directory location or a folder in Google Drive, Dropbox, etc. It tracks changes to files.\\
\cellcolor{gray!6}{GitHub} & \cellcolor{gray!6}{A remote commercial hosting service for Git repositories \citep{github-user-count}}\\
GitHub issues & A mechanism to track tasks or ideas\\
\cellcolor{gray!6}{commit} & \cellcolor{gray!6}{A set of saved changes to a local repo}\\
\addlinespace
pull & Update a local repo\\
\cellcolor{gray!6}{push} & \cellcolor{gray!6}{Upload local files to a remote repo}\\
forking & Create a copy of a repository under your account\\
\cellcolor{gray!6}{pull request} & \cellcolor{gray!6}{Propose changes to a remote repo}\\
merge conflict & Contradictory changes that cannot be integrated until they are reconciled by a user\\
\addlinespace
\cellcolor{gray!6}{branching} & \cellcolor{gray!6}{Keeping multiple snapshots of a repo}\\
gh-pages (GitHub Pages) & Special branch which allows creation of a webpage from within GitHub\\
\cellcolor{gray!6}{GitHub Actions} & \cellcolor{gray!6}{Mechanism for continuous integration}\\
GitHub Classroom & A system to facilitate distributing assignments to students. Instructors create a template Git repository that includes starter code, datasets, and document templates that students may need. A single URL is provided to the class, and each student is provided their own copy of the template repository when they click the URL and accept the assignment. The instructor can reuse the template repositories in future offerings \citep{github_classroom}.\\
\cellcolor{gray!6}{ghclass} & \cellcolor{gray!6}{An R package which provides an alternative system to GitHub Classroom to facilitate distributing assignments to students \citep{ghclass}}\\
\addlinespace
RStudio & An Integrated Development Environment (IDE), i.e., a front-end, for R that offers integration with Git. (\href{https://rstudio.com}{rstudio.com})\\
\cellcolor{gray!6}{RStudio Server Pro} & \cellcolor{gray!6}{A server-based version of RStudio that can be installed for free for academic use by instructors or institutions. (\href{https://rstudio.com/products/rstudio-server-pro}{rstudio.com/products/rstudio-server-pro})}\\
RStudio Cloud & A cloud-based version of RStudio software on servers provisioned by RStudio. (\href{https://rstudio.cloud}{rstudio.cloud})\\
\bottomrule
\end{longtable}

\hypertarget{motivation-for-version-control}{%
\subsection{Motivation for version
control}\label{motivation-for-version-control}}

There are two main motivations for including version control as a
learning objective in statistics courses. The first motivation is
reproducibility. For a scientific study to be replicated, the
statistical analysis in the study must be entirely reproducible.
Teaching reproducible analysis in the statistics curriculum helps make
students aware of the issue of scientific reproducibility and also
equips them with the knowledge and skills to conduct their future data
analyses reproducibly, whether as part of an academic research project
or in industry. \citet{baumer_2014} advocates teaching literate
programming early in the statistics curriculum via the use of R
Markdown, a system that enables students to produce computational
documents that includes their code, output, and written analysis using
the \texttt{rmarkdown} package \citep{rmarkdown_2018}. Literate
programming with R Markdown goes a long way towards computational
reproducibility, but a data analysis of considerable scope likely cannot
be managed in a single R Markdown document. As \citet{bryan_2018_excuse}
puts it, data analysis is an iterative process that relies on and
produces many files -- input data, source code, figures, tables,
reports, etc. Managing such projects is not unique to statistics, but it
is something that our curricula have been slow to address. Version
control provides a mechanism for managing all these files and sharing
them with others as a project progresses, and modern tooling and
workflows make it easier to implement in teaching than ever before.

The second motivation is industry and academic preparedness. The ability
to use version control systems is a highly desired skill in any industry
where writing code is part of the job, and the need to teach it has been
recognized in the literature \citep{haaranen_2015}. Git is a widely used
tool in industry for version control and code sharing. In a 2017 survey
of data scientists conducted by Kaggle, over 58\% of 6,000 survey
respondents remarked that Git was the main system used for version
control and code sharing in their workplace \citep{kaggle_2017}.
Additionally, knowing how to use GitHub is considered an essential skill
in the tech field, just as important as software development and
technical writing \citep{zagalsky_2015}. In an era where many of our
statistics and data science students are heading into jobs where they
will be writing code and working alongside software engineers and
developers, it is essential that we equip them with these skills.

Exposure to version control early (and often) in a statistics curriculum
ensures that, by the end, students not only enhance their statistical
analysis skills, but also develop workflows for conducting analyses
individually and collaboratively. With the widespread use of GitHub in
academia and industry, courses that teach version control prepare
students for internships, research programs, and their future careers.
More immediately, they can use these computing tools as they work on
analyses and projects in subsequent courses.

Additionally, implementing version control in a course can encourage
students to think about statistical analysis as an iterative process.
While working on a given assignment, students ``submit'' their work
multiple times by knitting their R Markdown file, writing a commit
message to document the changes, and pushing the updated work to their
assignment repository (repo). Some have adopted the mantra
``knit-commit-push'' for this workflow, and others ``commit-pull-push''.
Both are effective ways to help drive home the way that GitHub
structures and organizes changes to files.

Use of version control helps reinforce the notion that statistical
analysis typically requires multiple revisions, as students can review
their commits to see all the updates they've made to their work. A
desirable side effect is that, because students are periodically
``submitting'' their assignment as they work on it, there is less
pressure of the final deadline where everything must be submitted in its
final form. By the time a deadline approaches, students have ideally
submitted a majority of their work, which may reduce issues around late
submissions.

\hypertarget{method}{%
\section{Method}\label{method}}

In order to organize this paper, the authors first agreed upon a set of
organizing prompts to provide direction as they describe their
experiences:

\begin{itemize}
\tightlist
\item
  Describe the course/students.
\item
  Why use Git and GitHub?
\item
  What tools do you use for implementing Git in your class?
\item
  How do you introduce Git? Describe your students' first encounter with
  Git in the course.
\item
  What role does Git have in the regular day-to-day workflow for your
  students in the course?
\item
  How do you assess Git proficiency as a learning objective?
\item
  How do you address the United States Family Educational Rights and
  Privacy Act (FERPA) and related privacy issues?
\item
  Do you have other advice for instructors considering incorporating Git
  or some other type of version control into a statistics course?
\end{itemize}

The contributors were then free to address as many of these prompts as
they deemed appropriate. Each narrative description was then written
independently in an attempt to reduce cross-pollination and promote
similarities and differences to emerge naturally. The panel responses
have been organized to follow a similar structure within each section
(course description, tools and implementation, first exposure in class,
regular workflow, assessment, and other remarks). The order in which
these responses are presented aligns roughly with their place in a
statistics curriculum at each respective institution: a first year
undergraduate course, a second undergraduate course in statistics, a
course in a Master's in Statistical Science program, and finally a
Master's level course in a Biostatistics program.

\hypertarget{common-features-of-the-courses}{%
\section{Common features of the
courses}\label{common-features-of-the-courses}}

While taught at different levels and serving different audiences, there
are also a few aspects shared by all of the courses described. First,
all of these courses teach and use the R computing language along with
RStudio as the integrated development environment for R \citep{rstudio}.
Second, each course either requires or offers the option to access
RStudio in the browser, either using an RStudio Server Pro instance
hosted by their university or using RStudio Cloud, a cloud-based service
managed by RStudio \citep{rstudio_cloud}. Both options allow students
and the instructor to use the same versions of R, RStudio, and any
packages required for the course, which cuts down on early difficulties
related to managing local installations. Additionally, this means
students only need a web browser and an internet connection to access
the computing environment and can start programming as soon as the first
day of class \citep{cetinkaya_2018}. Nearly any computer, Chromebook, or
tablet is sufficient, and students can easily switch hardware as needed
if using one device (e.g., tablet or lab computer) in class and another
(e.g., personal computer) outside of class. In the rest of the paper we
will refer to these products generically as RStudio or the RStudio IDE.

\begin{figure}
\includegraphics[width=1\linewidth]{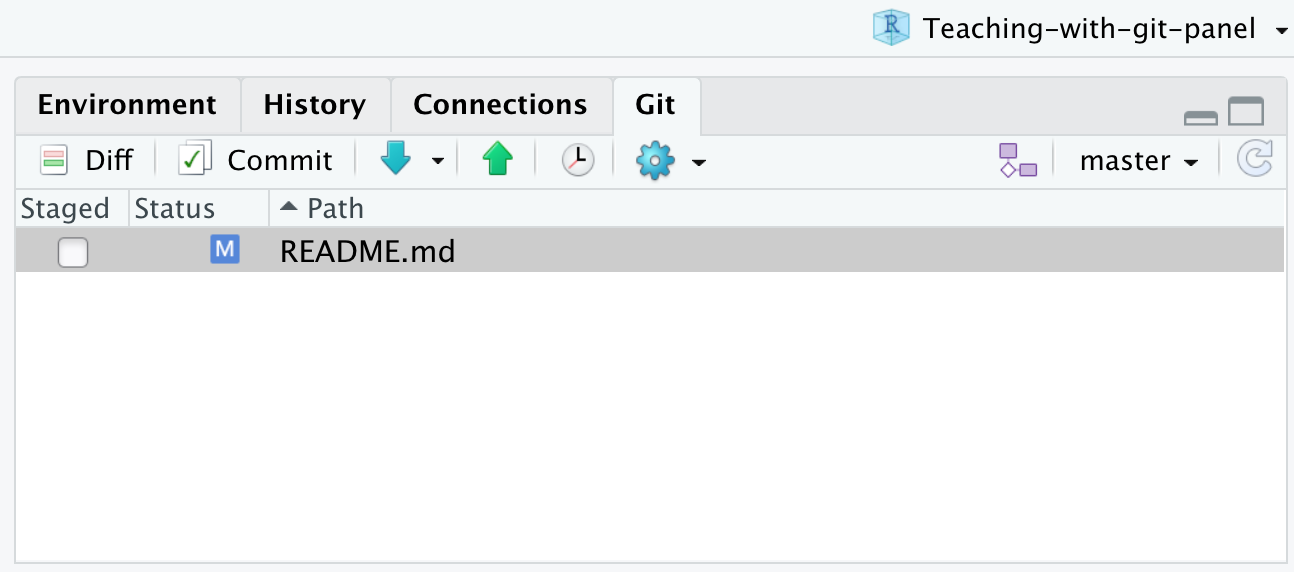} \caption{Example of Git pane within an RStudio IDE window}\label{fig:rstudio-git-pane}
\end{figure}

Server-based access to RStudio also streamlines Git installation and
integration with RStudio. Each course uses the Git pane in RStudio as
opposed to Git's command line interface or GitHub Desktop. The RStudio
interface is attractive since it is familiar to many students and it
facilitates use of the basic functionality of Git using a
point-and-click interface to practice version control fundamentals and
implement key steps of the workflow (e.g., \emph{diff}, \emph{commit},
\emph{pull}, \emph{push}; see Figure \ref{fig:rstudio-git-pane}). This
implementation serves to mitigate cognitive load while students gain
proficiency with unfamiliar tools and workflows, yet easily extends when
required, since the terminal is accessible within RStudio if shell
commands are necessary.

All of the instructors manage Git related course logistics by
establishing a GitHub Pro organization through GitHub Education. GitHub
provides unlimited free private repositories as well as compute-time
credits that can be used for running automated actions on the student
repositories. Use of private repositories ensures compliance with the
Federal Educational Rights and Privacy Act (FERPA) by protecting student
work and information from public access \citep{ferpa}. These private
repositories are only accessible to the student, the instructor, and any
other course teaching staff, e.g., teaching assistants. Additionally,
GitHub organizations hide the identity of all members to non-members by
default, meaning that students' enrollment in the course will not be
disclosed by their joining the course organization. Access within the
GitHub organization is further controlled on a per user basis via
permissions: the instructor and teaching assistants are ``Owners'' and
the students are ``Members''. As ``Owners'', the instructor and teaching
assistants are able to manage organization membership as well as create
and manage any repository created within the organization, and thus see
all students' work. As ``Members'', students are only able to view and
access the individual or team repositories assigned to them; they cannot
view or access the repositories for any other students. They, as well as
non-members, can also view any repositories within the organization that
have been made public, such as those containing supplemental notes or
any other materials for the course.

\hypertarget{first-year-data-computing-course}{%
\section{\texorpdfstring{First-year data computing course
(\Beckman{})}{First-year data computing course ()}}\label{first-year-data-computing-course}}

\hypertarget{course-description}{%
\subsection{Course description}\label{course-description}}

STAT 184: Introduction to R Programming at Penn State University is a
two-credit (2 x 50 minute instruction each week for 15 weeks) R
programming course originally modeled after a similar course and
accompanying textbook \citep{kaplan_2015, kaplan_beckman_2019} first
developed by Daniel Kaplan at Macalester College. This course is
designed for first-year undergraduate students from any academic program
and has no prerequisites. The course currently enrolls 30-40 students in
each of 9 sections per academic year, although enrollment demand has
been increasing rapidly since it was first developed in 2015. At least
one section each year is made available to first-semester students
interested in the statistics major (it has even been coordinated with
their orientation seminar class in the past) and at least one other
section is offered by popular demand to mixed audiences of any major and
class standing.

Major topics in STAT 184 include data wrangling and visualization with
\texttt{tidyverse} tools, literate programming with R Markdown, and
version control with Git. These themes persist for the entire semester
and are complemented by a survey of topics such as statistical
foundations, web scraping, regular expressions, simulation principles,
and basic machine learning ideas.

\hypertarget{tools-and-implementation}{%
\subsection{Tools and implementation}\label{tools-and-implementation}}

The workflow for students in STAT 184 includes the RStudio IDE, its Git
pane for version control, and interacting with GitHub. The only step in
the workflow that involves a tool outside of these is typically the Git
configuration step that students complete once at the beginning of the
semester using the Terminal pane of RStudio.

The instructors use one additional tool, GitHub Classroom
\citep{github_classroom}, to deploy assignments to students as described
in \citet{fiksel_2019}. GitHub Classroom facilitates the batch creation
of private student repositories on GitHub with starter documents for
each assignment such as instructions, a grading rubric, data sources,
and an R Markdown template.

Each week students are assigned one or more assignments with GitHub
Classroom. Students then clone these repositories, work on them and
commit and push their changes back to GitHub as they go, and finally
submit their assignments in the university's learning management system
(LMS). The end-of-semester project has a slightly different workflow;
work may be submitted as a GitHub repository or a website automatically
created using GitHub Pages, a setting configured within the project
repository.

\hypertarget{first-exposure-in-class}{%
\subsection{First exposure in class}\label{first-exposure-in-class}}

For the first student encounter with Git and GitHub in STAT 184, the
instructor creates a public GitHub repository associated with a GitHub
Pages website for the class (e.g., see
\href{https://mdbeckman.github.io/GitHub-Practice-StatChat-SP20}{mdbeckman.github.io/GitHub-Practice-StatChat-SP20}).
After a 15 minute class discussion to motivate version control as a
means to support collaboration and reproducibility as well as orient
students to a schematic representation of a workflow that includes Git
and GitHub, students are provided a link to the aforementioned GitHub
Pages website which includes instructions for a hands-on activity to be
completed during class. The activity walks through a first encounter
with Git and GitHub which invites students to (1) create a GitHub
profile of their own, (2) create their first Git repository, (3) turn
their personal repository into a GitHub Pages website, (4) edit a table
in the instructor's class repository to add their name, GitHub user ID,
and a functioning URL for GitHub Pages website they have just created,
(5) leave an informative commit message and initiate a \emph{pull
request} to the instructor's repository. The resulting table (from step
3) provides the instructor with the name and GitHub username associated
with each student, and every student creates a public GitHub Pages
website as a starting point to begin developing a work portfolio for
future use.

The exercise takes approximately 30 minutes and introduces key elements
of version control:

\begin{itemize}
\tightlist
\item
  Creating a repository.
\item
  Making a few commits and issuing a pull request.
\item
  Contributing to an outside repository belonging to another GitHub
  user.
\item
  Observing how to merge pull requests.
\item
  Observing how merge conflicts are created and resolved.
\end{itemize}

This short activity has the additional benefit that the entire exercise
can be completed within the GitHub web interface. The advantage of this
approach is that students do not need to use R, the Terminal, or even
Markdown, which allows them to begin building a schema for version
control without distraction from tools still unfamiliar to them at this
early stage.

\hypertarget{workflow}{%
\subsection{Workflow}\label{workflow}}

After the first exposure to version control as described above, each
assignment throughout the semester is associated with a private GitHub
repository that each student maintains. Nearly the entire workflow takes
place within the RStudio IDE. The only regular (albeit trivial) version
control task outside the RStudio IDE is the requirement that students
click a link to accept the template repository deployed through GitHub
Classroom, and then students are taken to a GitHub repository associated
with their personal copy of the assignment which they clone as an
RStudio project. From that point on, students can make a \emph{commit},
\emph{push}, \emph{pull}, view a \emph{diff} (difference between current
and previous versions of a document), etc. using the Git pane in
RStudio.

Some of these Git repositories are associated with an activity that is
launched, completed, and submitted in the space of a week or less, while
other Git repositories are used on a regular basis throughout nearly the
entire academic term. Additionally, a few repositories are associated
with collaborative assignments for which two or three students must
contribute by making commits to shared repositories. Students are
expected to make regular commits in each repository with the assessment
of several assignments taking into account their commit history evident
in the associated GitHub repos. However, the final product of most
assignments is submitted to the course LMS for grading.

\hypertarget{assessment}{%
\subsection{Assessment}\label{assessment}}

If version control is to be taken seriously as a learning objective for
a course, then it should be made clear to students early and often.
Students should find mention in the course syllabus, students should
expect to see it on exams, and students should feel that it is among the
class norms for regular assignments. In the earliest iterations of STAT
184 development, version control had been treated as an incidental
topic: encouraged, but not assessed. As a learning objective, version
control is now integrated into a wide variety of assessments, including
homework assignments, projects, and exams. For early assignments,
success is set at a relatively low threshold: objective evidence that
students have simply created a repository of their own or edited a
repository provided to them. As the workflow becomes more familiar,
assessment may include direct scrutiny of a commit history or similar
activity documented within a specific repository.

For example, students are expected to maintain a single repository for
all weekly problem sets assigned from the textbook during the semester.
This repository is then graded as a distinct homework score at the end
of the semester based on verification that all assignments are present
and associated with some minimum number of commits per assignment. To be
clear, the goal is simply to incentivize commits early and often in the
workflow of each assignment. Students should not be preoccupied by
counting commits; the actual number is inconsequential. Once version
control has truly taken root in the students' workflow, many STAT 184
students more than double the minimum number of commits required.

Lastly, students are to expect version control content on in-class
exams. For example, this might include open-ended tasks about important
concepts, selected-response questions (e.g., True/False or multiple
choice) about procedural details such as whether a \emph{pull} action
modifies files (a) in the directory on their local computer {[}correct
answer{]}, (b) on the GitHub Remote Server, (c) both, (d) neither, or
another task might prompt students to resolve an apparent merge conflict
presented to them in a screenshot.

\hypertarget{a-subsequent-data-science-course}{%
\subsection{A subsequent Data Science
course}\label{a-subsequent-data-science-course}}

STAT 184, along with an introductory statistics course (e.g., AP
Statistics), serves as a pre-requisite for an intermediate-level course
in the curriculum (STAT 380: Data Science Through Statistical Reasoning
and Computation) that is required for both statistics and data science
majors. This course extends practices, including version control,
introduced in STAT 184 such that the tools and workflow are largely
unchanged with the exception that students are expected to use a local
installation of RStudio. Since prior experience with version control in
the prerequisite STAT 184 class is assumed, STAT 380 instructors can
simply upload a roster of names and email addresses to GitHub Classroom
and students can identify themselves as they start their first
assignment. Assessments still include various version control elements
but are held to higher standards as expected of a more mature workflow.

\hypertarget{a-second-course-in-statistics}{%
\section{\texorpdfstring{A second course in statistics
(\Tackett{})}{A second course in statistics ()}}\label{a-second-course-in-statistics}}

\hypertarget{course-description-1}{%
\subsection{Course description}\label{course-description-1}}

STA 210: Regression Analysis is an intermediate-level course in the
Department of Statistical Science at Duke University. About 90 students
take the course each semester, representing a variety of majors across
campus. The course is one of the core requirements for the statistical
science major and minor, so a large proportion of the students intend to
pursue the major or minor. The only prerequisite is an introductory
statistics or probability course, therefore the students coming into STA
210 have a range of previous experiences using R and Git. As an example,
in Fall 2019, a majority of students had previously used R and RStudio
in another course, while less than half had any previous exposure to Git
and GitHub. Given the variability in previous experiences with these
computing tools, some of the challenges of teaching Git in this course
are similar to those experienced in a first semester statistics course.

\hypertarget{tools-and-implementation-1}{%
\subsection{Tools and implementation}\label{tools-and-implementation-1}}

The primary computing tools used in STA 210 are the RStudio IDE and
GitHub. On the instructor side, all administrative activities involving
GitHub are done using the \texttt{ghclass} R package \citep{ghclass}.
These activities include adding the students to the GitHub organization
at the beginning of the semester, creating teams on GitHub, making and
replicating assignment repos, and cloning the repositories for grading.
These processes are described in more detail in Section
\ref{tackett:workflow} and further details are provided in the
documentation for the \texttt{ghclass} package. The computing
infrastructure using RStudio and GitHub as well as the course pedagogy
is based on \citet{cetinkaya_2018} and Data Science in a Box
\citep{dsbox}.

\hypertarget{first-exposure-in-class-1}{%
\subsection{First exposure in class}\label{first-exposure-in-class-1}}

\label{tackett:first-exposure}

Students are introduced to Git at the very beginning of the semester. On
the first day of class, students create GitHub accounts with guidance
from \citet{bryan_2018_happy} on choosing a user name. At the beginning
of the semester, a portion of lecture is used to introduce version
control and reproducibility, why they are important, and how RStudio and
Git will help students implement these practices in their work.

One of the first assignments in the course is a computing assignment
focused on using RStudio and Git. This assignment serves as a review for
some students, and it is an introduction to these tools for others. For
all students, however, it is their first exposure to the workflow they
will use throughout the rest of the course. Students write their
responses in an R Markdown file, knit the file to produce a Markdown
document, write a short and informative commit message, and push their
work to GitHub for submission. Throughout the assignment instructions
are periodic reminders to \emph{knit}, \emph{commit}, and \emph{push}, a
mantra used throughout the semester to remind students how to connect
their work in RStudio to their assignment repository in GitHub. The
instructions for the first few assignments also include examples of
informative commit messages. The instructions end with a reminder for
students to review their work in the assignment repository on GitHub to
ensure it is the final version to be submitted for grading.

Students complete several assignments individually before working on
their first team assignment. This gives them an opportunity to become
familiar with RStudio and Git and become comfortable with this workflow
before introducing the additional layer of collaborating in GitHub. For
the first team assignment, in addition to the aforementioned workflow
cues, students also receive cues to \emph{pull} so they have the most
updated version of the collaborative document. There are also cues to
rotate which team member types the responses. These workflow cues are
eventually removed from the assignment instructions as the semester
progresses and the workflow is more routine for students.

\hypertarget{workflow-1}{%
\subsection{Workflow}\label{workflow-1}}

\label{tackett:workflow}

There are two basic workflows in the course: one for general
assignments, such as homework and computing labs, and one for the final
project.

The typical workflow for general assignments is the following:

\begin{enumerate}
\def\labelenumi{\arabic{enumi}.}
\tightlist
\item
  The instructor creates a starter repository. The starter repository
  includes a link to the assignment instructions, an R Markdown
  template, and a folder for the data (if needed). In the beginning of
  the semester, the dataset is already included in the starter
  repository. As the semester progresses, students download the data
  from the assignment instructions and upload it to the repository.
\item
  A copy of the starter repository is created for each student (or team)
  using the \texttt{ghclass} R package. For individual assignments, the
  repositories are named using the template
  \emph{assignment\_name-{[}user\_name{]}}, where
  \emph{{[}user\_name{]}} is the student's GitHub username. For team
  assignments, the repositories are named
  \emph{assignment\_name-{[}team\_name{]}}, where
  \emph{{[}team\_name{]}} is the team's name on GitHub. For example, the
  first individual homework assignment is named
  \emph{hw01-{[}user\_name{]}}.
\item
  Students start a new project in RStudio by cloning their assignment
  repository. They configure the RStudio project with the GitHub
  repository by using the \texttt{use\_git\_config()} function in the
  \texttt{usethis} R package \citep{usethis}. Students complete the
  assignments in RStudio, by typing their responses in an R Markdown
  document with \texttt{output:\ pdf\_document} which produces a PDF
  from the R Markdown document. They periodically \emph{knit},
  \emph{commit}, and \emph{push} their work to their repository on
  GitHub.
\item
  Students submit their work by connecting their repository to the
  associated assignment on Gradescope, an online rubric and grading
  system \citep{gradescope}.
\item
  Students view the assignment feedback on Gradescope. It is connected
  to the LMS which ensures that grades are securely stored within the
  university's system.
\end{enumerate}

During the second half of the semester, students complete a final
project in teams of three or four. The workflow for the project is
generally similar to the one described above, with the exception being
how students receive feedback. At various checkpoints in the project,
students receive feedback as an ``issue'' in the GitHub repository for
their project. They can reply to the issue as one way to ask the
instructor follow-up questions about the comments. This feedback
workflow is used for the project to more closely mimic how students may
exchange ideas with collaborators if they use GitHub outside of the
classroom setting. Though comments are posted in the GitHub issue, there
are no grades posted on GitHub. All grades related to the project are
posted only in the LMS.

\hypertarget{assessment-1}{%
\subsection{Assessment}\label{assessment-1}}

Developing a proficiency using RStudio and Git is a learning objective
for the course, so students are assessed on how they use the tools on a
majority of assignments. Students are required to have their work in
their GitHub repository to be considered for grading, so they must learn
how to \emph{push} to GitHub in order to complete individual assignments
and both \emph{push} and \emph{pull} for team assignments.

Each assignment includes a category named ``Overall'' that includes
points dedicated to using Git. Typically about 5\% of the points on an
assignment are for having at least three commits and writing informative
commit messages. On team assignments, there are also points allocated
for having at least one commit from each team member. This is used to
hold team members accountable for contributing and to encourage teams to
make use of the collaborative nature of GitHub.

\hypertarget{other-remarks}{%
\subsection{Other remarks}\label{other-remarks}}

Based on multiple semesters of teaching version control in undergraduate
courses, here are a few recommendations for instructors who are
considering teaching GitHub as a learning objective in an undergraduate
statistics course:

\begin{itemize}
\tightlist
\item
  Get early buy-in from students. As mentioned in Section
  \ref{tackett:first-exposure}, a portion of lecture in the beginning of
  the semester introduces the importance of reproducibility and version
  control. Given the relatively steep learning curve for Git and GitHub,
  it is important that students understand the value of learning these
  computing skills and how they are used when doing a statistical
  analysis.
\item
  Focus only on the GitHub functionally used in the course, as
  introducing too much functionality can become overwhelming. Generally
  teaching students how to \emph{push}, \emph{pull}, \emph{commit}, and
  resolve merge conflicts is enough to complete the assignments in a
  statistics course.
\item
  Utilize the Git pane in RStudio. Running Git commands through the
  RStudio interface helps make it more accessible for students who don't
  have previous experience running code from the command line.
\end{itemize}

\hypertarget{a-masters-level-statistical-programming-course}{%
\section{\texorpdfstring{A Master's level statistical programming course
(\Rundel{})}{A Master's level statistical programming course ()}}\label{a-masters-level-statistical-programming-course}}

\hypertarget{course-description-2}{%
\subsection{Course description}\label{course-description-2}}

STA 523: Statistical Computing was developed in 2016 for the (then) new
Master's in Statistical Science (MSS) program at Duke University and has
been offered yearly since. The course was designed around three core
pillars: focusing on reproducible methods, emphasizing programming
knowledge, and teaching foundational data science skills. The course
shares many commonalities with approaches to integrate computing
suggested by \citet{nolan_templelang_2010} (see
\url{https://www.stat.berkeley.edu/~statcur}).

The course is required for all first year MSS students in their first
semester, and consists of two 75 minutes lectures and one 75 minute
workshop per week. These students come from a wide variety of
backgrounds with many having little or no prior coding experience.
Similarly, most students have never used or have had minimal experience
with Git and GitHub or other version control systems. As the only
required course in the MSS program that focuses on computing,
programming, and software engineering, the goal of the course is to
provide the students with a strong foundation of skills that are
relevant to their other courses as well as their careers after
graduation. While the ideal computational skill set for an MS in
Statistics graduate is a moving target, it has become clear that beyond
traditional topics (e.g.~numerical computing, optimization, etc.) more
data-focused skills (e.g.~data munging, databases, SQL, etc.) are
increasingly important. We have attempted to reflect this in the
course's evolving curriculum. In addition, the course covers statistical
topics such as modeling and prediction as well as Bayesian methods such
as approximate Bayesian computation and MCMC. These statistical topics
are presented to complement other coursework in the curriculum by
focusing on the computational details and implementation.

\hypertarget{tools-and-implementation-2}{%
\subsection{Tools and implementation}\label{tools-and-implementation-2}}

Similar to the two previous courses, students in this course use the
RStudio IDE and interact with GitHub via the Git pane in RStudio. In the
earliest iterations of the course the process for creating,
distributing, and collecting student work from GitHub repositories was
done manually or via simple shell scripts. Over time a number of tools
have made this process much easier by allowing for the automation of
most of these processes. Specifically, GitHub released their Classroom
tool around the same time this course was first offered and it was used
for the delivery of individual assignments. However, its team assignment
workflow was initially not present and later overly constraining, as it
did not allow the instructor to assign teams.

These limitations and the power and availability of the GitHub API led
to the development of the \texttt{ghclass} R package which is used for
automating interactions with GitHub for course management
\citep{ghclass}. The package has more functionality than can be explored
in this paper, but the core use case is for automating the creation of
team and individual assignment repositories. For example, using the
template repository based workflow, described above, the package helps
create new repositories, create teams, add members, add teams or users
to the repositories, and then copy the template's contents with a single
function, e.g.~\texttt{org\_create\_assignment()}.

One additional attractive feature of the package is the ability to
interact with existing repositories, particularly when it comes to
adding or modifying files. This is valuable to address the
all-too-common situation that a distributed assignment includes a typo,
a minor issue with the data, or the repo contains the wrong version of a
file. Rather than having to send out an email announcing the issue or
posting an announcement to the course LMS, \texttt{ghclass} allows for
\emph{push}ing the corrected file(s) out to all of the students'
repositories in a way that is merged with any existing work. Since
everything is managed via Git there is no risk of, permanantly,
overwriting student work. The work that has gone into implementing the
necessary low level functionality in this package has allowed extensions
and experiments with the automation of higher level processes (such code
formatting feedback and peer review).

As some of the topics in the course involve fairly heavy computational
workloads, a centralized powerful departmental server hosting RStudio
Server Pro provides students access to the necessary compute resources.
The more extensive computational workload is also relevant to the usage
of large datasets in the course, as GitHub has a strict file size limit
of 100MB, while some of the datasets used in this class exceed multiple
gigabytes in size. To address this issue, these large files are hosted
in a shared, read-only directory on the same server that hosts RStudio,
such that all students have access to the files without having to
maintain their own copy. These data could also be hosted in the cloud,
but co-locating the data with the compute resources is important both
for efficiency and cost.

\hypertarget{first-exposure-in-class-2}{%
\subsection{First exposure in class}\label{first-exposure-in-class-2}}

As stated previously, the expectation is that students will use Git and
GitHub for all of their assignments within the class starting from day
one. In order to ease the initial learning curve for these tools, as
well as RStudio, the class explicitly includes at least one hour of
lecture in the first week to motivate these topics. Typically, this
takes the form of a very brief introduction to the theory and history of
version control and Git, and then the remainder of the time is dedicated
to a live demonstration of the tools.

At this time, the first assignment is distributed via a GitHub Classroom
link: students follow the link, connect their GitHub account to a unique
identifier in the roster, and then gain access to their own private
repository copy of the assignment template repository. As part of the
live demonstration, the instructor leads students through this process
and how to locate and interact with the repository on GitHub. This leads
to forking the repository, cloning the Git repository as an RStudio
project, and a demo of the basic usage of the Git pane. The basic Git
actions such as \emph{stage}, \emph{commit}, \emph{push}, and
\emph{pull} are covered and the class usually concludes by purposefully
inducing a merge conflict to demonstrate the process of resolving it.

This is a large amount of material for the students to absorb in a short
period of time. Recording the session (either via screen recording or
lecture capture), providing in-person support, and having an initial
individual assignment that is focused on reinforcing the workflow has
allowed most students to get up to speed quickly. The number of
Git-related questions during workshops and in office hours is
substantial during the initial weeks of each semester, but tends to
decrease rapidly as the semester progresses.

\hypertarget{workflow-2}{%
\subsection{Workflow}\label{workflow-2}}

The course features four to eight team assignments and two individual
projects, all of which require complete reproducibility for full credit.
For each assignment / project a template repository is created, which is
structured using an RStudio project and contains the files necessary for
the assignment. Typically, this template includes a README file which
contains a detailed description of the assignment, a scaffolded R
Markdown document which gives a uniform structure for the assignment and
includes clear indications of where the students should enter their
implementations / solutions and write up, and any additional necessary
support files (e.g., data, scripts, etc.). These template repositories
can easily be shared with teaching assistants for feedback and or
training purposes, and can also be moved from previous years into new
organizations for each new offering of a course. The work is assigned to
the students by mirroring the template repository to either the student
or their team's private repository, using a consistent naming scheme
e.g., \texttt{hw01-team01}, which gives them access to their own copy of
all the necessary files for the assignment and can be directly cloned as
an RStudio project from GitHub using the \texttt{New\ Project}
interface.

Students are then able to work on the assignment within RStudio and turn
in their work as well as collaborate with team members by
\emph{commit}ting and \emph{push}ing their code back to GitHub. Earlier
versions of the course focused on teaching both the RStudio Git
interface as well as the command line Git interface, but the vast
majority of students preferred the former so the latter approach was
dropped in favor of adding other content. At the deadline for each
assignment it is simply a matter of cloning all of the assignment
repositories from the organization to obtain a local copy of all of the
students' work, which can then be rerun to assess reproducibilty and the
resulting HTML or PDF documents graded.

\hypertarget{assessment-2}{%
\subsection{Assessment}\label{assessment-2}}

While grading each student or team's work, their R Markdown documents
are recompiled to ensure the reproducibility of their work. In early
versions of the course this was coupled with a course policy that work
that failed to compile would receive a zero, which turned out to be
almost impossible to enforce in practice. Most of the time students'
code would fail to run for relatively small issues (e.g., use of
\texttt{setwd()} to set working directories with absolute file paths or
loading a less commonly used package that the instructor did not have
installed) that could be easily fixed yet caused a compilation error. It
was possible to have a back and forth with the students about the errors
and have them correct them but this proved to be very inefficient and
frustrating for both the students and instructor.

The solution to this has been to implement automatic feedback for the
students on the basic processes of their assignment. This is done by
checking their R Markdown documents and repositories using continuous
integration tools available via GitHub Actions. These tests take the
student's code and run basic sanity checks every time students
\emph{push} their code to GitHub: does the R Markdown document knit, are
only the necessary files included in the repository? GitHub Actions are
used to implement these tests. The test results are signaled to students
via a badge in their repository README that shows either green or red
depending on whether the check passed or not. Additionally, they can
click on these badges to get specific feedback in the case a check
failed. This becomes a simple necessary (though not sufficient)
condition for the students to examine when completing an assignment.
Examples of the current GitHub action based workflows being used for
this and related courses are available at the \texttt{ghclass-actions}
GitHub repository \citep{ghclass_actions}.

Usage of Git and GitHub has never been an explicitly assessed component
of these courses, however it is inherently tied to the students work as
it is the only available method of obtaining and turning in their work.
There have not been any specific efforts to encourage particular
workflows with Git / GitHub (i.e., branching, issues, etc.) but it has
been interesting to observe some of the emergent behaviors that student
teams have developed for collaboration.

\hypertarget{other-remarks-1}{%
\subsection{Other remarks}\label{other-remarks-1}}

This specific course has been offered every Fall since 2016 and has
consistently had a capped enrollment of around 40 students per semester.
In 2017, it was decided to add an undergraduate equivalent to the
course, STA 323, which has been offered in the Spring semester each year
since and has also had a capped enrollment of approximately 40 per
semester. Lastly, upon joining the University of Edinburgh, a similar
Master's-level Statistical Programming course, Math 11176, has been
taught which has an enrollment of around 180 students.

The example of Jenny Bryan and her Stat 545 course \citep{bryan_stat545}
at the University of British Columbia served as a direct inspiration for
these courses. It was invaluable to see that such a course already
existed and had been successful over a number of years. The open
publishing and dissemination of the course materials directly influenced
many aspects of the class. Without this clear model of a course it is
unlikely that these courses would have been developed or been as
successful. More of this type of creative sharing would be of benefit to
the community.

\hypertarget{a-masters-level-biostatistics-course}{%
\section{\texorpdfstring{A Master's level biostatistics course
(\Sullivan{})}{A Master's level biostatistics course ()}}\label{a-masters-level-biostatistics-course}}

\hypertarget{course-description-3}{%
\subsection{Course description}\label{course-description-3}}

Statistical Programming in R (PHP 2560) is a first programming course
for Master's students in Biostatistics at Brown University. The main
focus of the course is to develop good statistical programming habits
while also preparing students for the world of reproducible research and
data science. The class started with 14 students and as of Fall 2019
drew more than 50 students a semester, including students from
undergraduate to PhD level in multiple departments across the
University. Many students come to the course with some experience using
R for data analysis, but no experience writing functions, designing
packages, or creating interactive web applications with the
\texttt{shiny} package \citep{shiny}.

\hypertarget{tools-and-implementation-3}{%
\subsection{Tools and implementation}\label{tools-and-implementation-3}}

Students use their own computers and have the choice between installing
both R and RStudio locally or via RStudio Cloud. Each student creates a
GitHub account and connects this to their RStudio use. Pre-class and
in-class coding exercises are all placed into GitHub repositories which
act as starter code for assignments created using GitHub Classroom. When
students accept the assignment, they use RStudio to clone their GitHub
repository as an RStudio project. They are guided to
\emph{commit-pull-push} frequently as they complete their work. As noted
earlier, GitHub Classroom provides a flexible platform for creating
private GitHub repositories for students.

\hypertarget{first-exposure-in-class-3}{%
\subsection{First exposure in class}\label{first-exposure-in-class-3}}

The course is about statistical programming in R, however, the first few
class meetings are dedicated to learning Git. Prior to any instruction
in R, the students work through the basics of Git. Before the first
class, students install Git, R and RStudio on their computer, then work
through the ``First days of Git'' Learning Lab
\citep{github_learning_lab}. During the first class, students follow a
link to a starter assignment from the course website and they create an
RStudio project with this link. Then, students work through a few
exercises throughout the entire first class practicing the Git workflow
and in RStudio.

\hypertarget{workflow-3}{%
\subsection{Workflow}\label{workflow-3}}

The rest of the course consists of DataCamp \citep{datacamp} assignments
and pre-class coding exercises. The pre-class work is all in one
repository to which students contribute throughout the semester. The
course website contains links to either GitHub repositories for students
to grab files from or an assignment link generated using GitHub
Classroom. The class consists of a 15 minute overview and review of
pre-class work. Then students begin a project by cloning their assigned
project in RStudio. Each class they first \emph{commit} their pre-class
work into the team repository. They then review and comment on each
others code and \emph{push} those comments and feedback. At this point,
they begin working on team coding projects. They typically create their
own scrap work file and then together decided on which code is displayed
on the teams' final results.

When they encounter their first merge conflict, they are instructed to
work through this with the help of the instructor as well as GitHub's
merge conflict tutorial \citep{github_merge_conflict}. The Git process
is slow going and some students find it to be frustrating at first, but
it does not take long before they have very few conflicts or problems
with their repositories.

\hypertarget{assessment-3}{%
\subsection{Assessment}\label{assessment-3}}

The instructor and teaching assistants create plain text files (e.g.,
\texttt{feedback.md}) in each repository and comment the code based on a
published rubric that all students see on the course site. Scoring
rubrics include evaluation of the repository history and commit messages
as well as R programming style. Feedback is then added to the student
repositories so they can \emph{pull} to see the remarks shared with
them.

\hypertarget{other-remarks-2}{%
\subsection{Other remarks}\label{other-remarks-2}}

The experience of teaching this course for over 4 years (6 offerings of
the course) led to the following recommendations for other instructors
considering teaching a course with version control as a learning
objective:

\begin{itemize}
\tightlist
\item
  Experiment with Git and RStudio implemented on various computers. If
  students run into problems while installing or configuring the tools,
  it is helpful to have experience with more than one implementation.
  Utilize students who have things working on their computer to help
  others troubleshoot. This creates teamwork but also allows for more
  students to receive support at a time.
\item
  Invest some time early to motivate the workflow with Git and address
  common pitfalls. For example, students should avoid committing files
  which they did not actually change to help minimize merge conflicts,
  and instead investigate why they show up in the Git pane in the first
  place. Each student working in shared repositories could be encouraged
  to maintain individual scrap work file in the repository that only
  they edit as a measure to avoid merge conflicts or overwriting someone
  else's work. Such files should be removed from the repository before
  final submission.
\item
  Be patient. Students and instructors alike may encounter challenges in
  the beginning, and sometimes it can be hard to diagnose the problem
  they are having. However, you will find their ability to code in teams
  and track their work begins to outweigh any issues.
\end{itemize}

\hypertarget{discussion-etinkaya-rundel-and-horton}{%
\section{\texorpdfstring{Discussion (\c{C}etinkaya-Rundel and
Horton)}{Discussion (etinkaya-Rundel and Horton)}}\label{discussion-etinkaya-rundel-and-horton}}

In addition to the organizing prompts, each contributor populated a
matrix of learning outcomes with codes representing the type of exposure
typical in the course(s) described. Table \ref{tab:learn-table} presents
this matrix using the following symbolic representation for each
learning outcome in each course:

\begin{itemize}
\tightlist
\item
  \(\square\square\square\square\): None. This is not included in
  course.
\item
  \(\blacksquare\square\square\square\): Incidental. This may or may not
  occur in the course.
\item
  \(\blacksquare\blacksquare\square\square\): Teacher. This is
  demonstrated to students, but they are not necessarily expected to do
  it independently.
\item
  \(\blacksquare\blacksquare\blacksquare\square\): Student. Students are
  expected to do this independently, but it may not be formally
  assessed.
\item
  \(\blacksquare\blacksquare\blacksquare\blacksquare\): Assessed.
  Students are expected to do this independently, AND will be assessed
  for proficiency.
\end{itemize}

\begin{table}[!h]

\caption{\label{tab:learn-table}Git learning outcomes and assessment across the courses.}
\centering
\begin{tabular}[t]{lcccc}
\toprule
Learning Outcome & \Beckman{} & \Tackett{} & \Rundel{} & \Sullivan{}\\
\midrule
\addlinespace[0.3em]
\multicolumn{5}{l}{\textbf{Repositories:}}\\
\hspace{1em}clone a private repository and push a commit & $\blacksquare\blacksquare\blacksquare\blacksquare$ & $\blacksquare\blacksquare\blacksquare\blacksquare$ & $\blacksquare\blacksquare\blacksquare\blacksquare$ & $\blacksquare\blacksquare\blacksquare\blacksquare$\\
\hspace{1em}create a repo & $\blacksquare\blacksquare\blacksquare\blacksquare$ & $\blacksquare\square\square\square$ & $\blacksquare\blacksquare\square\square$ & $\blacksquare\blacksquare\blacksquare\blacksquare$\\
\hspace{1em}create a branch, merge branches & $\square\square\square\square$ & $\square\square\square\square$ & $\blacksquare\blacksquare\square\square$ & $\blacksquare\blacksquare\blacksquare\blacksquare$\\
\hspace{1em}retrieve an older version of a file & $\blacksquare\blacksquare\square\square$ & $\blacksquare\blacksquare\square\square$ & $\blacksquare\blacksquare\square\square$ & $\blacksquare\blacksquare\blacksquare\square$\\
\hspace{1em}continuous integration or other automation & $\square\square\square\square$ & $\square\square\square\square$ & $\blacksquare\blacksquare\square\square$ & $\square\square\square\square$\\
\addlinespace[0.3em]
\multicolumn{5}{l}{\textbf{GitHub Issues:}}\\
\hspace{1em}create, comment on, and or assign an issue & $\blacksquare\blacksquare\square\square$ & $\blacksquare\square\square\square$ & $\blacksquare\square\square\square$ & $\blacksquare\blacksquare\blacksquare\square$\\
\hspace{1em}reference commits/code line numbers in issues & $\square\square\square\square$ & $\blacksquare\square\square\square$ & $\blacksquare\square\square\square$ & $\blacksquare\square\square\square$\\
\addlinespace[0.3em]
\multicolumn{5}{l}{\textbf{Collaboration:}}\\
\hspace{1em}student teams collaborate in a shared repo & $\blacksquare\blacksquare\blacksquare\square$ & $\blacksquare\blacksquare\blacksquare\blacksquare$ & $\blacksquare\blacksquare\blacksquare\blacksquare$ & $\blacksquare\blacksquare\blacksquare\blacksquare$\\
\hspace{1em}resolve a merge conflict & $\blacksquare\blacksquare\blacksquare\square$ & $\blacksquare\blacksquare\blacksquare\square$ & $\blacksquare\blacksquare\blacksquare\square$ & $\blacksquare\blacksquare\blacksquare\square$\\
\hspace{1em}fork and create a pull request & $\blacksquare\square\square\square$ & $\square\square\square\square$ & $\blacksquare\square\square\square$ & $\blacksquare\blacksquare\blacksquare\blacksquare$\\
\hspace{1em}merge a pull request & $\blacksquare\square\square\square$ & $\square\square\square\square$ & $\blacksquare\square\square\square$ & $\blacksquare\blacksquare\blacksquare\square$\\
\hspace{1em}review changes and blame & $\blacksquare\square\square\square$ & $\blacksquare\square\square\square$ & $\blacksquare\square\square\square$ & $\blacksquare\square\square\square$\\
\hspace{1em}create gh-pages & $\blacksquare\blacksquare\blacksquare\square$ & $\square\square\square\square$ & $\blacksquare\square\square\square$ & $\blacksquare\blacksquare\blacksquare\square$\\
\bottomrule
\end{tabular}
\end{table}

The instructors of the undergraduate courses had similar approaches to
the use of GitHub, with more advanced topics (e.g., branching and
continuous integration) only showing up in the graduate courses. There
was considerable heterogeneity between topics in terms of whether they
were formally assessed.

While there are a number of commonalities to these instructor stories,
there are also some differences. In this section we provide a high level
overview of some of the issues raised.

\hypertarget{students-need-to-see-value-of-these-expert-friendly-tools}{%
\subsection{Students need to see value of these expert-friendly
tools}\label{students-need-to-see-value-of-these-expert-friendly-tools}}

As instructors it can sometimes be frustrating to teach foundational
tools and approaches since students often want to jump directly to fancy
models or visualization. This may leave them unable to carry out simpler
and more straightforward tasks where their analyses can be documented
and reviewed.

How can we help to motivate students to think about the importance of
workflow and develop internal motivation? Peter Norvig (Google) notes
that what students need most is ``meticulous attention to detail''
\citep{nas_2019}. Are there ways that we can help them develop and
strengthen this capacity by demonstrating that source code control is a
tool that can be useful to tracking their work and to help them be less
error-prone? One approach might be to share a cautionary tale, perhaps
Xiao-Li Meng's story \citep{mengtimetravel} of the data loss of much of
his doctoral dissertation.

We saw multiple examples of instructor scaffolding to provide a guided
introduction to the power and value of GitHub (to track and document
their work) without getting lost in the details. We believe that the
scaffolded introduction to GitHub is a useful if not sufficient
framework to build more habits that foster reproducibility.

\hypertarget{start-slowly-and-keep-it-simple}{%
\subsection{Start slowly and keep it
simple}\label{start-slowly-and-keep-it-simple}}

A key takeaway of the approaches described in the paper are how
instructors have started slowly and gradually built up complexity. The
instructors adopted a ``less is more'' approach to avoid cognitive
overload. This is evidenced in how they each structure students' first
exposure to version control as well as how they (almost) all limit
interactions with version control to a small number of Git actions
through the RStudio IDE. While courses that feature teamwork get into
thorny concepts like merge conflicts, these are deferred until later in
the course, after students develop more comfort managing version
control.

Other more advanced features of Git (e.g., branching, pull requests,
rebasing, HEAD) are both valuable and commonly used by data scientists,
but such details might be appropriate to leave until later. Having
students learn to use straightforward Git workflows early on is a big
step on their path to developing good habits for workflow and
collaboration.

We should also note that keeping it simple doesn't necessarily mean that
students won't learn about more advanced Git tricks from other
resources. For example, it is possible to backdate a Git commit. Since
there is no notion of a ``deadline'' in Git repositories, students could
presumably backdate a commit made after the deadline for an assignment
as though it was made before the deadline. It is, however, possible to
prevent students from making changes to their repositories after a
deadline via a few indirect methods. Instructors can collect (clone or
download the contents of) student repositories at the deadline.
Alternatively, instructors can change permission levels of students at
the deadline so that they can no longer \emph{push} changes, but can
continue to read and interact with the repository. Both of these methods
can be automated using the \texttt{ghclass} package.

\hypertarget{why-github-and-not-gitlab-bitbucket-etc.}{%
\subsection{Why GitHub (and not GitLab, Bitbucket,
etc.)?}\label{why-github-and-not-gitlab-bitbucket-etc.}}

There are a number of web-hosting platforms for projects version
controlled with Git. The three most popular of these platforms are
GitHub, GitLab, and Bitbucket. Among these, GitHub is recognized as the
industry standard platform for hosting and collaborating on version
controlled files via Git with an estimated more than 2.1 million
businesses and organizations using GitHub \citep{github-user-count},
compared to an estimated number of one million users on Bitbucket
\citep{bitbucket-user-count} and more than 100,000 organizations on
GitLab \citep{github-user-count}.

In addition, GitHub provides a rich API, which allows for tools like
GitHub Classroom and \texttt{ghclass}. The \texttt{ghclass} package
offers functionality for peer review by moving files around between
GitHub repositories of students. Additionally, features like GitHub
Actions can be used for immediate feedback and auto-grading of files by
triggering certain code to run in the background every time students
\emph{push} to their repositories.

One potential difficulty with using GitHub is the fact that the student
code---even in private repositories---is hosted on GitHub servers, which
means student data leaves the university. This is especially important
for institutions outside of the US since there may be laws around
student data leaving the country and being stored on US servers, e.g.,
the European Union's General Data Protection Regulation (GDPR). One
remedy to this is for the university to enter a data protection
agreement with GitHub for GDPR compliance. Another possibility is for
the university to host their own GitHub server. The software associated
with this, GitHub Enterprise, is freely available for academic teaching
use. However the university needs to supply the hardware (server) as
well as IT resources to set up the server and student authentication.
Note that the latter is a much more resource intensive solution.

We note an important potential danger of building curricula solely
around any specific commerical technology, including GitHub. Until 2018,
GitHub was a start-up. That year it was acquired by Microsoft. It's
impossible to tell what is next for the company. The company currently
seems dedicated to continue offering free private organizations and
repositories for educational use, and we have no reason to doubt this.
However, it serves as a reminder that software tools and their terms of
use do change from time to time. It is certainly a risk, but we note
that, if you're teaching data science, and want to stay current,
instructors should be willing to take a certain amount of risk, in a
calculated way such that the students don't end up suffering
consequences harshly. Many educators and developers are investing time
in building infrastructure and tooling to help others stay current with
their data science pedagogy and tooling. Instructors, even those not
interested in participating in the development of such tools, should
track what is being developed to ensure their programs stay current in
this rapidly changing environment.

\hypertarget{not-one-single-path}{%
\subsection{Not one single path}\label{not-one-single-path}}

One striking take-home observation from the instructors' stories is that
different instructional teams follow different models with different
pedagogical goals, ranging from fostering collaboration or automating
aspects of course mechanics. Some of the ways students and instructors
use GitHub include: (1) a pull request model, (2) full write access to
individual or team project repos, (3) use of the \texttt{ghclass}
package with one repository per student per assignment, or (4) use of
GitHub Classroom with one repository per student per assignment.

There are also various approaches to assessing student work and
providing feedback on GitHub. Use of issues to provide feedback is a
popular approach, and it is possible to make this process more efficient
and streamlined with the use of issue templates. Automated feedback
using continuous integration tools like GitHub Actions is another
approach that can either replace or supplement the manual feedback
process via issues. Writing automated tests to fully evaluate data
science assessments that not only contain code and output but also
interpretations is difficult, and perhaps impossible. Therefore, it's
difficult to imagine how automated checks can fully replace manual
grading, but development on this front are exciting to see as statistics
and data science classes grow in size.

A key implication of these various approaches is that we don't want to
be too prescriptive in terms of a specific workflow. We might consider
an analogy to code style: there are multiple reasonable ones, and we can
(and do) engage in somewhat religious arguments about what is ``right''
but the key aspect is that there are compelling reasons to ``fit in''.
The same approach is important when thinking about teaching GitHub and
version control.

Some of the structures described in the paper (e.g., the
\texttt{ghclass} package), are powerful and flexible systems that
facilitate scaling to larger classes. As a reviewer notes, these systems
have a non-trivial learning curve. For classes with no team-based work,
and especially for an instructor who is just starting with Git and
GitHub, we recommend using GitHub Classroom for managing the
distribution and collection of students assignments as repositories. For
classes that also involve teamwork, the \texttt{ghclass} package offers
more complete functionality for course management. Additionally, the
\texttt{ghclass} package also provides support for automation of editing
and correcting repositories that have already been distributed to
students as well as automating many other common tasks that need to be
applied across a large number of repositories via the GitHub API, e.g.,
managing organization and team membership, retrieving repository
statistics, testing using GitHub Actions, etc. As of the time of writing
this paper, these features are not offered in GitHub Classroom. It
should also be noted that GitHub Classroom and the \texttt{ghclass}
package are compatible tools and can be used in alongside each other.

\hypertarget{peer-review}{%
\subsection{Peer review}\label{peer-review}}

Another approach that several of us are exploring in our courses is peer
review, which has the benefit of exposing students to each others' work
and also prepares them for industry settings where code review is
commonplace. GitHub is already designed for peer review as this is a
crucial part of software development. From technical perspective, peer
review is enabled with the functions starting with the \texttt{peer\_*}
prefix in the \texttt{ghclass} package. They offer functionality for
retrieving files from one repository, anonymizing by stripping the
metadata (e.g., student names, commit history), moving these files to a
new repository where a randomly selected student has access to read and
review, and collecting these reviews and submitting them as a pull
request to the original student repositories. The full peer review
functionality and process is described in detail in a vignette in the
\texttt{ghclass} package \citep{ghclass}. Peer review gives students the
opportunity to meaningfully engage with each others' work and learn from
each other. It also gives them a chance to try to reproduce others' work
and experience the difficulties of reproducing others' work.

But the workflow that is native to GitHub (via branches, pull requests,
and no anonymity) does not always work for teaching -- either because
some of these concepts are beyond the learning goals of the course
(e.g.~in intro courses we don't talk about branches and pull requests)
or may not be suitable for a learning environment (an instructor might
want to do anonymous peer review so students don't know whose work
they're reviewing).

GitHub's rich API allows us to leverage what's already built in to
GitHub and customize it to be more suitable for peer review as part of
coursework. Several of the authors are exploring how best to integrate
peer review facilitated by GitHub into our courses.

\hypertarget{creating-portfolios}{%
\subsection{Creating portfolios}\label{creating-portfolios}}

Students sometimes use their GitHub profile as part of their job search
\citep{portfolio}. Educators may encourage their students to curate
their GitHub profile based on their coursework. Use of GitHub in courses
may assist with this process. However this is not automatic as
coursework needs to be stored in private repositories and it's not
always obvious how to expose this work while conforming to FERPA, GDPR,
etc.

One approach is to only allow students to convert their final project
repository to be public, and let them know of specific assignments from
class that are approved for reproducing in public repositories. These
are usually assignments with low risk of plagiarism and/or high
personalization. Moreover, it is important for instructors to keep
summative or overly critical feedback and any grades out of repositories
that students might convert to public repositories later. Any team work
also needs to be handled with care: all team members need to agree that
work can be made public.

It should also be noted that it can take time for students to get their
class projects into something that is portfolio worthy. Often times, a
repository with just some code doesn't make a compelling portfolio
entry. Students need to create an informative but brief write-up that
features highlights from their work, so that those browsing their
portfolio know where to start looking, or more importantly, why this
repository is worth looking into. One quick solution for this is a rich
README. An additional step is to publish the repository as a webpage
simply by turning on GitHub Pages (gh-pages) feature, which will turn
the README of a repository into a webpage. This process can be an
official part of an assignment or provided to students as a parting gift
to help increase their visibility on the web. Future research in this
area might explore ways that such e-portfolios might be helpful in both
curating student work and highlighting their efforts.

\hypertarget{assessment-4}{%
\subsection{Assessment}\label{assessment-4}}

Many of the instructors have touched on the commit history providing a
transparent account of the work done by students in their repositories,
which can be especially useful for individual accountability in
teamwork. While number of commits, on its own, is not a strong indicator
of the quality of work done by a student, lack of commits can signal
that the student has not made direct contributions to a team project.
However, unless the course makes it clear that commits by each team
member are required, this number alone might not be a true
representation of a student's contribution. For example, if pair
programming without switching roles, commits would appear to be made
only by one student. We recommend making use of peer evaluations in any
courses that involve teamwork in order to get a clearer picture of each
students' contribution, and supplementing the feedback from these
evaluations with commit history statistics.

Since version control is a learning goal for these courses, we believe
that instructors should assess how well students are following
recommended workflows in each assignment. We recommend assigning roughly
10\% of the points in each assignment to organization and style (e.g.,
figure sizing, code style, formatting, etc.), and portion of those
points specifically to version control related tasks. These include: (1)
reasonable number of commits, (2) reasonable commit content, and (3)
meaningful commit messages. Getting statistics like number of commits is
made possible with the \texttt{ghclass} package. Assessing reasonable
commit content can be a lot more cumbersome, and likely only worth
looking into if the student has too few commits on an assignment.
Finally, for assessing whether the commit messages are meaningful, we
recommend quickly taking a peek at the commit history on the GitHub
repository, and scanning to see if there are any commit messages that
don't obviously meet this criteria (e.g., random string of characters or
too many commits that just say ``update'').

It is crucial for instructors to model good version control hygiene
before assessing it, as students can't be expected to come into the
class with an intuition for what is reasonable commit content or
message. One way of doing this is explicitly stating when and what to
commit in earlier assignments (e.g., ``make a commit after this
exercise'') and what to say in the commit message. Throughout the
semester this sort of scaffolding can be slowly removed from
assignments, letting the students learn to make a decisions about what
constitutes a reasonable change that should be captured in a commit.
(It's also imperative that the instructor practice what they preach in
terms of instructor commits in course and student repositories.)

\hypertarget{automation-and-workflow}{%
\subsection{Automation and workflow}\label{automation-and-workflow}}

More advanced users (and many instructors) will benefit from the use of
automation tools. For the instructor, this might facilitate auto-pulling
local files, simplify returning feedback to students (e.g., a script to
open all repositories to the issues page, a script to pull files, add a
file called \texttt{feedback.md} that includes comments, \emph{commit},
and \emph{push} to many repos). The use of continuous integration tools
to check for compilation may be particularly helpful as students work on
more complex tools and approaches that go beyond the capabilities of
their laptops. This is an area where new approaches are being developed
in the research community that have the potential to improve student
experiences and/or simplify work by the instructor and improve learning
outcomes for students.\\
We hope that this paper encourages instructors to explore and share
their experiences.

\hypertarget{closing-thoughts}{%
\subsection{Closing thoughts}\label{closing-thoughts}}

Students heading into the workforce need to be able to structure,
organize, and communicate their work. Using version control is a
valuable, useful, and now logistically practical tool, so we recommend
instructors consider incorporating it into their courses and programs.

It's worth mentioning that there are plenty of viable alternatives to
the RStudio IDE (e.g., Atom, JupyterHub, Vim) or Git for version control
(e.g., Subversion, Mercurial). Several of these tools have similarly
efficient integration, and it should be clear that no attempt was made
to compare and contrast alternative implementations in this paper.

One limitation of this article is that it only provides the educator's
point of view to using Git and GitHub. We have observed a few patterns
emerging in how students work with version control and how they
collaborate over a version control platform. Students tend to figure out
the basics of working with version control on individual assignments
pretty quickly, within the span of a few weeks. However they often find
collaboration, and especially merge conflicts, more challenging. To help
ease this challenge students may get together in person for team
projects so that they can make commits from a single computer, which can
also be seen a positive outcome, but it shows that students find
collaborating on GitHub challenging. Additionally, whether there is long
term adoption of using version control by the students is less clear.
However we observe that students coming out of these courses and then
working research projects or participating in ASA DataFest, a
weekend-long, team-based data analysis competition \citep{gould_2014},
often choose to use version control and collaboration with Git and
GitHub.

Finally, version control has not come up as an issue students bring up
in course evaluations, in fact, it regularly gets mentioned positively
among skills they learned in the courses. None of the classes mentioned
in this paper have systematically collected data on student attitudes
towards version control. We believe that this would be a valuable next
step for statistics and data science courses so that we can explore how
the implementation of GitHub in the classroom is associated with
students' classroom experiences, similar to \citet{hsing_2019}, which
discusses such a study conducted on students in computer science
courses. Such assessment data and other findings informed by the
learning sciences would help improve instruction in this area. Future
research could help inform a publication akin to Hesterberg's
\citeyearpar{hesterberg_2015} guide to teaching resampling entitled
``What every statistics and data science instructor should know about
version control and reproducible workflows''.

\pagebreak

\bibliographystyle{agsm}
\bibliography{bibliography.bib}

@article{mengtimetravel,
  title = {{XL-Files}: Time Travel and Dark Data},
  author = {Xiao-Li Meng},
  journal = {IMS Bulletin},
  volume = 49,
  number = 1,
  pages = {6},
  howpublished = {\url{https://imstat.org/wp-content/uploads/2019/12/Bulletin49_1.pdf}},
  year = {2020}
}

@article{nolan_templelang_2010,
  title = {Computing in the statistics curriculum},
  author = {Deboran Nolan and Duncan {Temple Lang}},
  journal = {The American Statistician},
  volume = 64,
  number = 2,
  pages = {97--107},
  year = {2010}
}

@misc{asa_guidelines_2014,
  title = "Curriculum guidelines for undergraduate programs in statistical science",
  url = {http://www.amstat.org/education/curriculumguidelines.cfm},
  author = {{American Statistical Association}},
  year = 2014,
  note = {Accessed: 2020-06-07}
}

@book{nasem_2018, 
	author = {{National Academies of Science, Engineering, and Medicine}},
	title = {Data Science for Undergraduates: Opportunities and Options},
	year = {2018},
	url = {https://nas.edu/envisioningds},
  	note = {Accessed: 2020-06-07}
}

@article{fiksel_2019,
  title = {Using {GitHub} classroom to teach statistics},
  author = {Jacob Fiksel and Leah R. Jager and Johanna S. Hardin and Margaret A. Taub},
  journal = {Journal of Statistics Education},
  volume = 27,
  number = 2,
  pages = {100--119},
  year = {2019}
}

@article{garfield_2011,
  title = {Rethinking assessment of student learning in statistics course},
  author = {Joan Garfield and Andrew Zieffler and Daniel Kaplan and George W. Cobb and Beth L. Chance and John P. Holcomb},
  journal = {The American Statistician},
  volume = 65,
  number = 1,
  pages = {1--10},
  year = {2011}
}

@article{hardin_2015,
  title = {Data science in statistics curricula: Preparing students to 'Think with data'},
  author = {Johanna S. Hardin and Roger Hoerl and Nicholas J. Horton and Deborah Nolan and Ben Baumer and Olaf Hall-Holt and Paul Murrell and Roger D. Peng and Paul Roback and Duncan Temple Lang and Mark D. Ward},
  journal = {The American Statistician},
  volume = 69,
  number = 4,
  pages = {343--353},
  year = {2015}
}

@article{bryan_2018_excuse,
  title = {Excuse me, do you have a moment to talk about version control?},
  author = {Bryan, Jennifer},
  journal = {The American Statistician},
  volume = {72},
  number = {1},
  pages = {20--27},
  year = {2018},
  publisher = {Taylor \& Francis}
}

@article{baumer_2014,
  title = {R Markdown: Integrating a reproducible analysis tool into introductory statistics},
  author = {Baumer, Ben and Cetinkaya-Rundel, Mine and Bray, Andrew and Loi, Linda and Horton, Nicholas J},
  journal = {Technology Innovations in Statistics Education},
  volume = {8},
  number = {1},
  year = {2014},
  url = {https://escholarship.org/uc/item/90b2f5xh},
}

@Book{rmarkdown_2018,
    title = {R Markdown: The Definitive Guide},
    author = {Yihui Xie and J.J. Allaire and Garrett Grolemund},
    publisher = {Chapman and Hall/CRC},
    address = {Boca Raton, Florida},
    year = {2018},
    url = {https://bookdown.org/yihui/rmarkdown},
  }

@inproceedings{haaranen_2015,
	author = {Haaranen, Lassi and Lehtinen, Teemu},
	title = {Teaching Git on the Side: Version Control System As a Course Platform},
	booktitle = {Proceedings of the 2015 ACM Conference on Innovation and Technology in Computer Science Education},
	series = {ITiCSE '15},
	year = {2015},
	isbn = {978-1-4503-3440-2},
	location = {Vilnius, Lithuania},
	pages = {87--92},
	numpages = {6},
	url = {http://doi.acm.org/10.1145/2729094.2742608},
	doi = {10.1145/2729094.2742608},
	acmid = {2742608},
	publisher = {ACM},
	address = {New York, NY, USA},
}

@misc{kaggle_2017, 
	title = {Kaggle Machine Learning \& Data Science Survey 2017}, 
	url = {https://www.kaggle.com/kaggle/kaggle-survey-2017}, 
	author = {Kaggle}, 
	year = {2017}, 
	month = {Oct}
}

@inproceedings{zagalsky_2015,
	author = {Zagalsky, Alexey and Feliciano, Joseph and Storey, Margaret-Anne and Zhao, Yiyun and Wang, Weiliang},
	title = {The Emergence of {GitHub} As a Collaborative Platform for Education},
	booktitle = {Proceedings of the 18th ACM Conference on Computer Supported Cooperative Work \& Social Computing},
	series = {CSCW '15},
	year = {2015},
	isbn = {978-1-4503-2922-4},
	location = {Vancouver, BC, Canada},
	pages = {1906--1917},
	numpages = {12},
	url = {http://doi.acm.org/10.1145/2675133.2675284},
	doi = {10.1145/2675133.2675284},
	acmid = {2675284},
	publisher = {ACM},
	address = {New York, NY, USA},
}

@Manual{rstudio, 
  title = {RStudio: Integrated Development Environment for R}, 
  author = {{RStudio Team}}, 
  organization = {RStudio, PBC.}, 
  address = {Boston, MA}, 
  year = {2015}, 
  url = {http://www.rstudio.com}, 
  note = {Accessed: 2020-06-07}
  }

@article{cetinkaya_2018,
	author = {Mine {\c{C}}etinkaya{-}Rundel and Colin Rundel},
	title = {Infrastructure and Tools for Teaching Computing Throughout the Statistical Curriculum},
	journal = {The American Statistician},
	volume = {72},
	number = {1},
	pages = {58--65},
	year  = {2018},
	publisher = {Taylor & Francis},
	doi = {10.1080/00031305.2017.1397549},
	URL = {https://doi.org/10.1080/00031305.2017.1397549},
	eprint = {https://doi.org/10.1080/00031305.2017.1397549}
}

@misc{portfolio,
	title = {What do job-seeking developers need in their {GitHub}?},
	author = {{Tech Beacon}},
	URL = {https://techbeacon.com/app-dev-testing/what-do-job-seeking-developers-need-their-github},
	year = 2020,
	note = {Accessed: 2020-06-07}
}

@misc{ferpa,
	title = {Federal Educational Rights and Privacy Act (FERPA)},
	URL = {https://www2.ed.gov/policy/gen/guid/fpco/ferpa/index.html},
	year = 2020,
	note = {Accessed: 2020-06-07}
}

@book{kaplan_2015, 
	author = {Kaplan, Daniel T}, 
	title = {Data Computing: An introduction to wrangling and visualization with R}, 
	year = {2015},
	publisher = {Project Mosaic}
}

@book{kaplan_beckman_2019, 
	author = {Kaplan, Daniel T. and Beckman, Matthew D.},
	title = {Data Computing}, 
	edition = {2}, 
	year = {2019},
	url = {https://dtkaplan.github.io/DataComputingEbook}
}

@misc{github_classroom, 
  title = {{GitHub Classroom}}, 
  author = {{GitHub Education}},
  url = {https://classroom.github.com}, 
  year = 2020,
  publisher = {GitHub},
  note = {Accessed: 2020-06-07}
}

@misc{ghclass,
	author = {Colin Rundel and Mine {\c{C}}etinkaya{-}Rundel and Therese Anders},
	title = {ghclass: tools for managing classes with {GitHub}},
	URL = {http://github.com/rundel/ghclass},
	year = {2020},
  	note = {Accessed: 2020-06-07}
}

@Manual{rstudio_cloud,
  title = {RStudio Cloud},
  author = {{RStudio Team}}, 
  organization = {RStudio, PBC.}, 
  address = {Boston, MA}, 
  year = 2020,
  url = {https://rstudio.cloud},
  note = {Accessed: 2020-06-07}
}

@misc{dsbox,
	title = {Data Science in a Box},
	author = {Mine {\c{C}etinkaya-Rundel}},
	URL = {https://www.datasciencebox.org},
	year = 2020,
	note = {Accessed: 2020-06-07}
}

@book{bryan_2018_happy,
  title = {Happy Git and GitHub for the useR},
  author = {Bryan, Jennifer},
  year = {2018},
  url = {https://happygitwithr.com},
  publisher = {GitHub},
  note = {Accessed: 2020-06-07}
}

@misc{usethis,
	author = {Hadley Wickham and Jenny Bryan},
	title = {usethis: Automate Package and Project Setup},
	URL = {https://github.com/r-lib/usethis},
	year = {2020},
	note = {Accessed: 2020-06-07}
}

@misc{gradescope,
	title = {Gradescope},
	URL = {https://www.gradescope.com},
	year = 2020,
	note = {Accessed: 2020-06-07}
}

@misc{ghclass_actions, 
	title = {ghclass actions}, 
	url = {https://github.com/rundel/ghclass-actions}, 
	author = {Colin Rundel}, 
	year = {2020}, 
	month = {Jan},
  	note = {Accessed: 2020-06-07}
}

@misc{bryan_stat545,
	title = {STAT 545 Website},
	author = {Bryan, Jennifer},
	URL = {https://stat545.com},
	year = 2020,
	note = {Accessed: 2020-06-07}
}

@Manual{shiny,
  title = {shiny: Web Application Framework for R},
  author = {Winston Chang and Joe Cheng and JJ Allaire and Yihui Xie and Jonathan McPherson},
  year = {2019},
  note = {R package version 1.4.0},
  url = {https://CRAN.R-project.org/package=shiny},
}

@misc{datacamp,
	title = {DataCamp},
	URL = {https://datacamp.com},
	year = 2019,
	note = {Accessed: 2019-12-19}
}

@misc{github_learning_lab,
	title = {{GitHub Learning Lab}},
	author = {GitHub},
	URL = {https://lab.github.com},
	year = 2020,
	note = {Accessed: 2020-06-07}
}

@misc{github_merge_conflict, 
	title = {{GitHub Learning Lab}},
	author = {GitHub},
	URL = {https://lab.github.com/githubtraining/managing-merge-conflicts},
	year = 2018,
	note = {Accessed: 2020-06-07}
}

@misc{bitbucket-user-count,
	title = {Bitbucket},
	URL = {https://bitbucket.org},
	year = 2020,
	note = {Accessed: 2020-06-07}
}

@misc{github-user-count,
	title = {GitHub},
	URL = {https://github.com},
	year = 2020,
	note = {Accessed: 2020-06-07}
}

@misc{nas_2019,
  title = "Roundtable on Data Science Postsecondary Education Meeting 10",
  url = {https://nas.edu/dsert},
  author = {{National Academies of Science, Engineering, and Medicine}},
  month = "March",
  year = 2019,
  note = {Accessed: 2020-06-07}
}

@incollection{gould_2014,
  title = {Teaching statistical thinking in the data deluge},
  author = {Gould, Robert and {\c{C}}etinkaya-Rundel, Mine},
  booktitle = {Mit Werkzeugen Mathematik und Stochastik lernen--Using Tools for Learning Mathematics and Statistics},
  pages = {377--391},
  year = {2014},
  publisher = {Springer}
}

@inproceedings{hsing_2019,
 author = {Hsing, Courtney and Gennarelli, Vanessa},
 title = {Using {GitHub} in the Classroom Predicts Student Learning Outcomes and Classroom Experiences: Findings from a Survey of Students and Teachers},
 booktitle = {Proceedings of the 50th ACM Technical Symposium on Computer Science Education},
 series = {SIGCSE '19},
 year = {2019},
 isbn = {978-1-4503-5890-3},
 location = {Minneapolis, MN, USA},
 pages = {672--678},
 numpages = {7},
 url = {http://doi.acm.org/10.1145/3287324.3287460},
 doi = {10.1145/3287324.3287460},
 acmid = {3287460},
 publisher = {ACM},
 address = {New York, NY, USA},
}

@article{hesterberg_2015,
  title = {What teachers should know about the bootstrap: resampling in the undergraduate statistics curriculum},
  author = {Hesterberg, Tim},
  journal = {The American Statistician},
  volume = 69,
  number = 4,
  pages = {371--386},
  year = {2015}
}

\end{document}